\documentclass[journal]{modcls}                     %
\onlineid{0000}

\usepackage{amssymb}

\long\def\ignore#1{}

\title{VizCV: AI-assisted visualization of researchers' publications tracks}

\author{%
  Vladimír Lazárik,
  \authororcid{Marco Agus}{0000-0003-2752-3525},
  \authororcid{Barbora Kozlíková}{0000-0003-0045-0872}, and 
  \authororcid{Pere-Pau Vázquez}{0000-0003-4638-4065}
}

\authorfooter{
  \item
  	Vladimír Lazárik is with Masaryk University.
  	E-mail: vladimir.lazarik@gmail.com.
  \item
  	Marco Agus is with Hamad Bin Kalifa University.
  	E-mail: magus@hbku.edu.qa.
\item
  	Barbora Kozlíková is with Masaryk University.
  	E-mail: kozlikova@fi.muni.cz.

  \item Pere-Pau Vázquez is with Universitat Politècnica de Catalunya.
  	E-mail: pere.pau.vazquez@upc.edu.
}

\abstract{%
    Analyzing how the publication records of scientists and research groups have evolved over the years is crucial for assessing their expertise since it can support the management of academic environments by assisting with career planning and evaluation. We introduce \textit{VizCV}, a novel web-based end-to-end visual analytics framework that enables the interactive exploration of researchers' scientific trajectories. It incorporates AI-assisted analysis and supports automated reporting of career evolution. Our system aims to model career progression through three key dimensions: \emph{a)} research topic evolution to detect and visualize shifts in scholarly focus over time, \emph{b)} publication record and the corresponding impact,  \emph{c)} collaboration dynamics depicting the growth and transformation of a researcher's co-authorship network. AI-driven insights provide automated explanations of career transitions, detecting significant shifts in research direction, impact surges, or collaboration expansions. The system also supports comparative analysis between researchers, allowing users to compare topic trajectories and impact growth. Our interactive, multi-tab and multiview system allows for the exploratory analysis of career milestones under different perspectives, such as the most impactful articles, emerging research themes, or obtaining a detailed analysis of the contribution of the researcher in a subfield. The key contributions include AI/ML techniques for: \emph{a)} topic analysis, \emph{b)} dimensionality reduction for visualizing patterns and trends, \emph{c)} the interactive creation of textual descriptions of facets of data through configurable prompt generation and large language models, that include key indicators, to help understanding the career development of individuals or groups. We demonstrate our approach with case studies of established researchers' profiles, obtained through Scopus. These case studies demonstrate how AI-assisted visual analytics facilitates more in-depth insights into academic careers. %

}


\teaser{
  \centering
  \includegraphics[width=0.9\linewidth, alt={Three views showing the evolution of the researcher, as a vertical dot timeline, a scatterplot of all the publications, and a description of the main insights of the researcher.}]{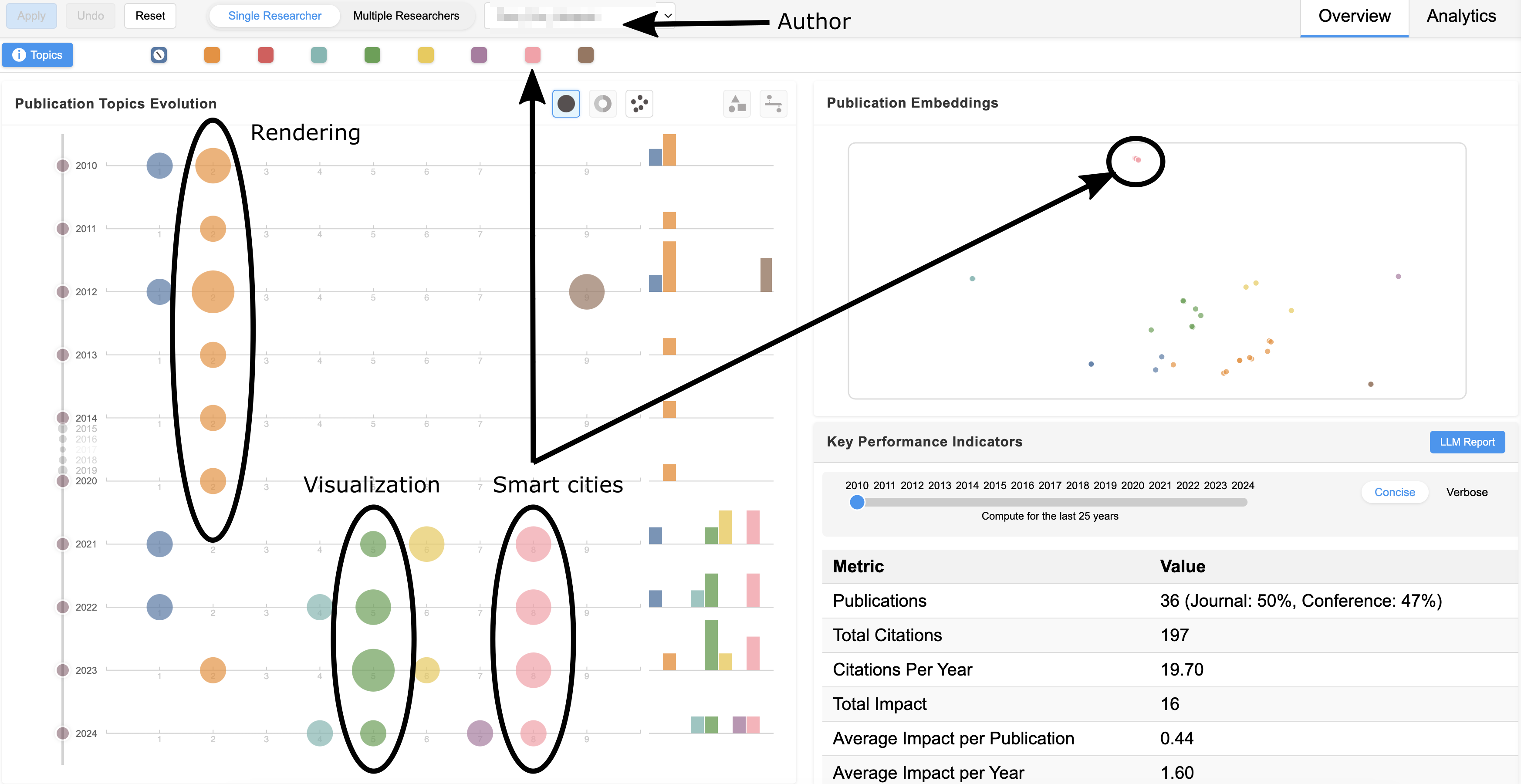}
  \caption{%
  \textbf{VizCV dashboard:}
   we show the initial view of our application comparing between two selected time periods (2010-2014 and 2020-2024). The left view shows the evolution of the researcher's topics of interest during these periods. It is visible that this researcher is now working in areas related to Visualization (green) and Smart Cities (pink), which were not present in the 2010-2014 period, where their main interest was mostly Rendering and Computer Graphics (orange). The rightmost view shows how this new area seems to be far from the previous topics (top group of pink dots). The bottom-right view shows a summary of key indicators and can be used to automatically generate a descriptive LLM report of the scientist that analyzes the indicators selected by the user. %
  }
  \label{fig:teaser}
}

\graphicspath{{figs/}{figures/}{pictures/}{images/}{./}} %

\usepackage{tabu}                      %
\usepackage{booktabs}                  %
\usepackage{lipsum}                    %
\usepackage{mwe}                       %

\usepackage{mathptmx}                  %

\begin{document}

\firstsection{Introduction}

\maketitle

In the academic environment, the examination of publication track records is commonly required to address various tasks, such as assessing the suitability of a researcher for a role in a commission, seeking an expert for collaboration, the evaluation of performance, or for recruitment purposes, to name a few. However, analyzing scientific careers may be a very time-consuming and challenging process, and many factors contribute to its complexity. For example, researchers may change their primary focus and topics of interest over the years. In some cases, the change can be significant. An exemplary case is Francis Crick, who started as a physicist, then shifted to biology, and ultimately was one of the proponents of the DNA helix structure~\cite{Watson_Crick}. Understanding how a scientist has evolved is not obvious when just skimming through the list of publications that can be obtained at popular databases such as Google Scholar, Web of Science, or Scopus, especially for very productive and long-established researchers~\cite{martin2018google}. Therefore, an end-to-end tool where the user can upload the profile of one or multiple researchers and obtain actionable insights on the career of the researcher, together with a report that summarizes the analysis of the desired metrics, would be of great help for many academic tasks. 

Although substantial research has already been devoted to analyzing collaboration or citation networks~(see Sec.~\ref{sec:related}), the examination of the evolution of research careers has attracted much less attention, and tools that analyze the evolution of topics of interest are limited in their features. VisPUBCompas \cite{Wang_2019}, for instance, analyzes Visualization publications at the institutional level. Despite the publications of individual authors being displayed, no elements are provided to analyze the researcher's evolution, such as the impact of the publications, and no direct comparison is offered either. Other systems concentrate on the impact of the list of papers through citations and collaborations and neither analyze the subfields of interest, nor enable direct comparisons~\cite{Wang_2018}. And descriptive tools, such as Vis Author Profiles~\cite{Latif_2019}, provide the users with a set of indicators in a textual description, through the aid of embedded charts. However, no direct comparison is available and the amount of information displayed is quite limited. Moreover, its database contains Visualization researchers, and the user cannot upload the desired profiles. To the best of our knowledge, no current tool provides a holistic representation and exploration of a researcher's publication track over the years, encompassing and combining topic analysis, impact, co-authorship evolutions over time, and researchers' comparisons.

To fill these gaps, we propose a novel end-to-end framework dubbed \textit{VizCV}, composed of a data processing workflow and a web-based visual analytics tool to display and analyze a researcher's career evolution. The tool enables users to solve problems such as finding adequate reviewers for articles, instructors for advanced courses, or tenure evaluation. 
Additionally, our decision-support system can generate automated AI-based personalized textual reports highlighting user-selected key performance indicators. This allows for a direct comparison of two or more scientists’ track records and topics of interest, as well as the evolution of the impact of their publications. To summarize, our contributions are:

\begin{itemize}
    \item We propose an end-to-end workflow to transform a (set of) Scopus profile(s), into a set of topics, scientometrics, and other derived data, to populate a set of views that enable the exploration and comparison of researchers' publication tracks. 
    \item An interactive multiview, web-based visual analytics tool that enables the examination of the evolution of topics and other scientometrics from individual researchers and the direct comparison between multiple researchers.
    \item A module to generate reports of researchers by generating and executing prompts to a Large Language Model (LLM) that analyzes relevant insights from a researcher. The output is a textual description highlighting the selected key outcomes of a researcher's career or a subset of their publications.
    \item An interactive visual prompt generator that lets the user select the key performance indicators of a researcher, their aggregation factor, and the type of report, and generates a prompt injected with the relevant data, ready to use with any LLM. 
\end{itemize}

Our proposed tool focuses on three distinct areas of researchers' interest, which address diverse aspects of their publication track:
\begin{itemize}
    \item \textit{General information:} This includes information about individual publications and their evolution over time.
    \item \textit{Publications' insights:} Topics and areas the researcher has published in, their evolution over time, and the number of co-authors for each publication.
    \item \textit{Metrics:} Measures that answer questions such as how two different authors have contributed to a certain field, scientometrics of the publications track records, such as the impact, citations, etc.
\end{itemize}

We demonstrate the capabilities of 
\textit{VizCV} on various analysis cases related to real-world application scenarios, and we 
report on a qualitative performance assessment.

\section{Related Work}
\label{sec:related}

Our work deals with comparative visual analytics applied to scientific publication records of scholars. The common procedure for the analysis of dissemination outputs, routinely performed by academic professionals and managers for quantitative assessment, consists of accessing large-scale publication databases, such as Google Scholar, Elsevier’s Scopus, and Clarivate’s Web of Science~\cite{mikki2009google} to collect metadata and citation information that support common metrics like the h-index, Impact Factor, and various journal rankings~\cite{martin2018google}. These bibliometric indices help evaluate the reach and influence of authors’ work, identify top venues, and track collaborations. However, while such metrics offer high-level summaries, they provide static representations and do not allow for dynamic and interactive visual analysis.
For this reason, over the past decade, visualization research has proposed interactive systems to explore an individual's publication record, the topics they study over time, their citation impact, and even to compare multiple researchers. These solutions often combine temporal visualizations (timelines, streams, evolving networks) with interactive filtering to help users trace the evolution of research focus, collaboration networks, and citation influence year by year. Below, we review the recent approaches falling into the following categories:  (1) visualization of academic and bibliographic data, (2) visual analytics for scholarly exploration, and (3) scientometric and egocentric visualization.

\paragraph*{Visualization of Academic and Bibliographic Data}
Interactive visualization frameworks for bibliographic data have been extensively explored to facilitate literature analysis, bibliometric evaluation, and trend discovery~\cite{Fung_2016, Mechant_2014, Shi_2015, Wu_2016}. Approaches such as topic evolution visualization~\cite{Dang_2019, He_2016}, egocentric networks~\cite{Zhao_2016, Shi_2015, Wang_2022}, and citation pattern analysis~\cite{Heimerl_2016, Nakazawa_2018} highlight varied methods for understanding scholarly impact. 
More specifically, Dang et al.~\cite{Dang_2019} presented WordStream, an interactive visualization technique that combines a streamgraph with a word cloud to show the evolution of topics of interest over time. It was used to visualize topical trends in document collections, including academic corpora. 
Fung et al.~\cite{Fung_2016} analyzed three alternatives for visualizing an individual’s publication record: a node-link diagram, an adjacency matrix, and a tree-based metaphor. The tree represents each publication as a leaf node, although it is slightly unclear what branches mean. It seems that they either group work by topic or venue. The goal was to provide a visual summary of one researcher’s output and how it clusters by research area. The authors found the tree metaphor effective for highlighting an individual’s research topics hierarchy and temporal output. 
Wang et al.~\cite{Wang_2022} focused on the influence of individual and social factors elements in the success of researchers' careers. To do so, they first leveraged a multifactor impact analysis framework to determine the effect of different elements on academic career success over time. Then, they created a multiview visualization tool that lets the user analyze three concepts: factors, categories, and the person. 
Very recent works enriched interactive systems with advanced embedding techniques to improve analytical capabilities~\cite{beck:2025:pure, qiu:2025:vadis, hao:2023:ieeevis, hoque:2024:dataopsy, huang:2023:vaembstar}.
For a general overview of methods integrating embeddings in visual analytics, we refer readers to the recent survey from Huang et al.~\cite{huang:2023:vaembstar}.
For specific bibliometric analysis, PURESuggest~\cite{beck:2025:pure}  combines citation-based suggestions with augmented visualizations of the citation network, while VADIS~\cite{qiu:2025:vadis} introduces a prompt-based attention model (PAM) that generates dynamic document embedding and document relevance adjusted to the user's query.

\begin{table*}[htbp]
    \centering
    \caption{Comparative analysis of selected bibliographic visualization systems.}
    \label{tab:comparative}
    \resizebox{\textwidth}{!}{%
    \begin{tabular}{lcccccc}
        \toprule
        \textbf{System} & \textbf{Year} & \textbf{Egocentric} & \textbf{Semantic Embeddings} & \textbf{Topic Evolution} & \textbf{Temporal Comparison} & \textbf{Topic Comparison} \\
        \midrule
        WordStream \cite{Dang_2019} & 2019 &  &  & \checkmark &  &  \\
        NLP Scholar \cite{Mohammad:2020:nlp} & 2020 &  &  & \checkmark &  &  \\
        Influence Flowers \cite{Shin_2019} & 2019 & \checkmark &  &  & \checkmark &  \\
        GeneticFlow \cite{Xiao_2023} & 2023 & \checkmark & \checkmark &  &  &  \\
        LitSense \cite{Sultanum_2020} & 2020 &  & \checkmark &  &  & \checkmark \\
        VISPubComPAS \cite{Wang_2019} & 2019 &  &  & \checkmark &  & \checkmark \\
        ACSeeker \cite{Wang_2022} & 2022 & \checkmark &  & \checkmark & \checkmark &  \\
        PUREsuggest \cite{beck:2025:pure} & 2025 &  & \checkmark &  & \checkmark &  \\
        VADIS \cite{qiu:2025:vadis} & 2025 &  & \checkmark & \checkmark &  &  \\
        \midrule
        \textbf{VizCV (ours)} & 2025 & \checkmark & \checkmark & \checkmark & \checkmark & \checkmark \\
        \bottomrule
    \end{tabular}}
\end{table*}

\paragraph*{Visual Analytics Systems for Scholarly Exploration}
Interactive visual analytics frameworks such as VISPubComPAS~\cite{Wang_2019}, LitSense ~\cite{Sultanum_2020}, and ACSeeker~\cite{Wang:2021:success} provide environments for exploring scholarly datasets with a particular focus on user-guided exploration, faceted filtering, and comparison. 
More recently, VA systems have started integrating semantic embeddings and transformer-based models~\cite{witschard:2025:visextr, qiu:2024:docflow}, enhancing user interaction and interpretability.
For example, Witschard et al.~\cite{witschard:2025:visextr} proposed a prevalence-aware method for topic extraction, and showed how it can be used to obtain important insights from two text corpora with very different content. Whereas DocFlow~\cite{qiu:2024:docflow} incorporates a query-based retrieval model that prioritizes documents with a high likelihood of answering the user's query.

\paragraph*{Scientometric and Egocentric Visualizations}
Recent contributions have addressed author-centric exploration, capturing patterns like self-citation, gender disparities, and interdisciplinary collaborations~\cite{Milz:2018:citations, Mishra_2018, Sarvghad_2022, Gomez:2022:leading, Mitrović_2023}. Approaches such as GeneticFlow~\cite{Xiao_2023, Luo_2023}, Influence Flowers~\cite{Shin_2019}, and SD$^2$~\cite{Guo_2022} introduced novel representations of academic influence and scholar profiling.
For example, Wang et al. proposed ImpactVis~\cite{Wang_2018}, an interactive visualization tool for exploring a researcher’s impact through citation data. The tool fills a gap between simple metrics (like h-index) and complex network analysis by providing an intuitive interface to probe citation patterns.
Shin et al.~\cite{Shin_2019} introduced a visual metaphor, called “Influence Flower”, to depict the bi-directional citation influence between a target researcher (the “ego”) and other entities. An Influence Flower is essentially a radial visualization: the ego is at the center, and the “petals” around are other authors or topics that either influenced the ego or were influenced by them.
GeneticFlow, developed by Xiao et al.~\cite{Xiao_2023}, is a scholar-centric visualization grounded in the concept of a self-citation network. It combines topic evolution with citation impact in one view: the self-citation graph inherently shows the evolution of the scholar’s ideas, while topic coloring and impact charts contextualize this evolution in terms of content and recognition.
Other systems have also centered on collaboration networks as part of academic career visuals. For instance, ResumeVis~\cite{zhang:2018:resumevis} incorporated an ego-network view to summarize a person’s co-working relationships, and ACSeeker~\cite{Wang:2021:success} includes a measure of collaboration diversity. 
Compared to existing interactive methods, our proposed framework uniquely combines egocentric exploration, advanced semantic embeddings, interactive visualization of citation patterns, and dynamic topic evolution analysis within a single integrated pipeline. In Table~\ref{tab:comparative}, we overview the key features of recent methods and show how they focus on limited aspects of scholarly data. For instance, GeneticFlow~\cite{Xiao_2023} emphasizes scholar-centric profiling but lacks a dynamic visualization of topic evolution. Conversely, PUREsuggest~\cite{beck:2025:pure} offers robust embedding-based document retrieval without providing egocentric citation dynamics. Our approach bridges these gaps by integrating comprehensive exploration, comparison capabilities, and semantic insights into a cohesive and AI-assisted interactive user experience.

\section{Analyzing researchers' publication records}

Two of the authors have extensive academic experience, and have been in different management roles at universities and external evaluation commissions. Therefore, they have been exposed to many of the consideration and decision procedures commonly used to understand, assess, evaluate, and compare the scientific performance of researchers. For example, topics of interest are regularly examined when filling evaluation boards, and impact metrics are commonly used in recruiting faculty procedures. Furthermore, in order to evaluate the performance of individual researchers, metrics such as citations, number of contributions to journals, and international collaboration networks are analyzed and eventually compared with other candidates, usually for periods of 5 to 10 years. 

As said, lists of articles, such as those obtained through Google Scholar, are insufficient to obtain detailed insights into the researcher's topics of interest and their evolution. Here, we demonstrate that many relevant insights can be extracted from information that is easily available, such as Scopus profiles.

\subsection{Data Scope and Acquisition}
Our system is intended for use in any research field. Therefore, we aim to include information covering various aspects of publications, such as the authorship of the publication, title, year, citation information (e.g., the citation count), impact, DOI, and content information, such as the abstract and keywords. This information can be obtained through Scopus in a double-step process. The first step is to extract the general information using the Scopus API or the website. Then, for each publication, the impact factor and relative impact factors can be automatically obtained using the Scopus API by a custom script. We collect both the Source Normalized Impact per Paper (SNIP) and the SCImago Journal Rank (SJR)~\cite{Roldan:2019:Current}. SNIP measures the average citation impact of a journal's publications by correcting for differences in citation practices between scientific fields. SJR is a measure of scientific influence that measures both the number of citations received by a journal and the importance or prestige of the journals from which they are coming. Even though rankings are prone to all sorts of non-ethical practices (e.g., journals may induce submitting authors to cite papers published in the journal)~\cite{Joshi:2024:Deception}, universities still use them extensively for evaluation and recruitment processes. Our goal is to demonstrate how they can be integrated into a visual analytics tool to better understand a researcher's profile. Scopus impact factors and citations can easily be interchanged with Google Scholar or Web of Science indicators.

\subsection{Requirements}
Based on our previous experience in several academic committees and managerial positions, we have identified a set of requirements our system should solve that belong to three different areas: General Information (GI), Publication Insights and Metrics (PIM), and Comparison (CMP).
General information about a researcher refers to a high-level overview of a person's publication track record, including:
\begin{itemize}
    \item \textbf{GI1:} Articles that have been published, with access to the details of their title, authors, venue, year, $\ldots$ 
    \item \textbf{GI2:} Topics of research and their evolution over time.
    \item \textbf{GI3:} Details of the publications (authors, abstracts, venue, $\ldots$).
    \item \textbf{GI4:} Descriptive overview of a researcher's career.
    \item \textbf{GI5:} Number of articles over the years.
\end{itemize}

\begin{figure*}[ht]
    \centering
    \includegraphics[width=0.7\linewidth]{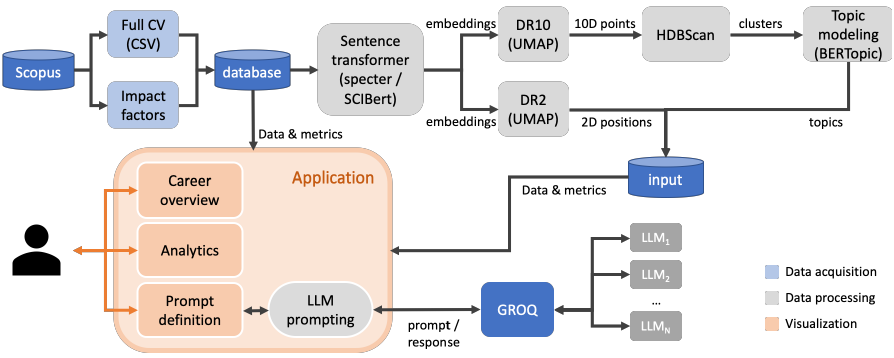}
    \caption{Pipeline of our data processing stack. The Scopus abstracts and keywords are processed through Sentence Transformer, to obtain embeddings for the documents. These embeddings are reduced to 2D for the scatterplot, and to 10D to create clusters using HDBSCAN. Topics are then extracted using BERTopic.}
    \label{Fig: Pipeline1}
    
\end{figure*}

For a more detailed examination of a researcher's track, the system should provide additional insights into the scholarly activities of one or multiple researchers. The following elements are commonly employed for assessment, promotion, and evaluation processes at universities: 

\begin{itemize}
    
    \item \textbf{PIM1:} Number of papers in each topic and how they evolve.
    \item \textbf{PIM2:} Number of authors per publication and how it evolves.
    \item \textbf{PIM3:} Impact of the publications (absolute and relative) according to Scopus metrics.
    \item \textbf{PIM4:} Evolution of the impact relative to other metrics, such as the number of authors.
    \item \textbf{PIM5:} Descriptive overview of author's articles (or the papers) from a given period.
\end{itemize}

Finally, for supporting other comparisons, such as when evaluating a person (or multiple candidates) for a committee or position, it is very helpful to have a set of views that can facilitate the comparative tasks: 

\begin{itemize}

    \item \textbf{CMP1:} Comparative evolution of topics per author.
    \item \textbf{CMP2:} Comparative evolution of citations for multiple authors.
    \item \textbf{CMP3:} Comparative evolution of the relative impact of multiple authors.
\end{itemize}

Although some of these requirements can be partially addressed using other existing tools, they do not provide a comprehensive and complete view of a researcher's publication tracks and their comparison. Many of these tools do not facilitate understanding the changes in topics of a researcher along their careers, or do not enable contextualizing a certain publication within the whole. For example, how impactful is a researcher in one area, or how many collaborators they have and whether they come from the same institution, or other countries. 

These requirements led us to create a multi-tab application that has a specific focus on the evolution of topics, publications, citations, as well as other parameters, in the course of a researcher's career.

\section{Application architecture}
\label{sec: Application architecture}

The data necessary to populate the different views was obtained through the end-to-end pipeline illustrated in Figure~\ref{Fig: Pipeline1}. The input is a database we created by first downloading the Scopus profiles and augmenting them with the SJR and SNIP impact factors of the publications through a Python script, and the citation data with another. The data is then processed using the following steps:

\begin{itemize}
    \item \textbf{Embedding generation:} The abstracts and keywords of each publication are joined and embedded separately using Specter~\cite{Cohan:2020:Specter}, a transformer-based system that has been trained using citations to inform the model. 
    \item \textbf{Topic detection:} Topics are detected using the multistep pipeline of the BERTopic algorithm \cite{Grootendorst:2022:Bertopic} with minor modifications to re-classify the detected topic classes:
    \begin{enumerate}
        \item The embeddings are reduced to 10D using UMAP~\cite{McInnes:2018:UMAP}, 
        \item then the clustering using HDBSCAN is applied~\cite{McInnes:2017:HDBSCAN},
        \item and finally, topics are detected and refined using a class-based variation of TF-IDF~\cite{Grootendorst:2022:Bertopic} (c-TF-IDF). We then apply a reclassification algorithm (explained later). %
    \end{enumerate}
            \item \textbf{Layout generation:} The 2D layout is built using UMAP to reduce the dimensions of the embeddings to 2D, and colored with the topics detected in the previous step. 
    \item \textbf{Generation of the data:} The resulting data, referred to as \emph{input}, is then used by our application as the data source.
\end{itemize}

Our visual analytics application then reads the information from the processed input and creates a web-based dashboard that has two different configurable tabs that let the user access the general information, analytics, and make multiple comparisons among different profiles. Both views include a set of key performance indicators and the option to configure and compose the textual survey with our report generator.

\paragraph*{Topic Detection}
One crucial aspect of the tool is the selection of topics. The quality of topic generation depends on multiple parameters, such as the dimensions used for the dimensionality reduction algorithm or the parameters of the HDBSCAN algorithm. We conducted a systematic evaluation process to balance the number of topics and the number of unclassified articles with 27 researchers from different areas, resulting in a total of 3887 unique papers after erasing duplicates.
For the dimensionality reduction, we ran HDBSCAN on embeddings reduced using UMAP with 2, 5, 10, 20, 50, and 100 dimensions. The 10D embedding represents a good trade-off between the number of unclassified documents and the size of the topics. We used a \emph{min\_cluster\_size} value of 7, \emph{min\_samples} value of 4, and an \emph{epsilon} value of 0.2. To reduce the number of small clusters, we designed the following algorithm: 
\begin{enumerate}
    \item Select all small topic clusters (size $<$ 40).
    \item Extract the topic embeddings from the BERTopic and compute cosine similarities between them.
    \item Iteratively re-classify individual small clusters into the most similar larger ones. We opted for custom modification of this step, instead of using the built-in mechanisms of BERTopic because the number of topics required by the algorithm cannot be easily known beforehand, and the automatic topic reduction performed by the algorithm was merging clusters based on highest similarity and not preserving the big and middle-sized clusters, instead of just re-classifying the small ones.
\end{enumerate}

The topic names generated by the c-TF-IDF are fine-tuned using a sequence of two algorithms provided by the BERTopic library: \emph{KeyBERTInspired} and \emph{MaximalMarginalRelevance}. The first method assures a closer relationship between the keywords and the actual contents of the papers assigned to the topic, by taking the most relevant documents and using them as the updated topic embedding. The second step reduces redundancy of keywords and enhances diversity by discarding keywords that already have similar alternatives in the representation.

\section{Analyzing the publication topics and insights}
\label{sec: Analyzing the publication interests}

The tabs of our proposed tool allow the user to focus on diverse aspects of the researchers' careers. The first tab, illustrated in Figure~\ref{fig:teaser}, provides an overview of a single researcher's publications, grouped by topics, or a comparative visualization of the contributions of two different scientists to the topics.
The second tab is called \emph{Analytics} and provides insights into selected scientometrics of the individual researcher. By selecting the option \emph{Multiple Researchers} in this tab, the system will enable a comparison view that provides a time-based comparison of the scientometrics of multiple researchers at once.

\subsection{Evolution and Comparison of Research Topics}

Experts searching for collaborators or university officials seeking a suitable person, e.g., for a committee, need to identify researchers with publications in a certain area. A list of publications that can be obtained on a platform like Google Scholar or from a CV does not easily communicate in which fields the author primarily publishes, in what quantities, or whether their research interest has shifted to another area over time. Although partial information can be extracted from the titles of the publications in university databases or services such as Google Scholar, these may not be informative enough for people outside that particular area. 
Thus, making sense of an author's expertise based on information extracted in this way may become an extremely cumbersome and time-consuming endeavor. 

To better understand the evolution of research topics over the course of a researcher's career, our application should be able to help solve the following tasks:

\begin{itemize}
    \item \textbf{T1:} Determine the topics of interest and how they have evolved over the years.
    \item \textbf{T2:} Identify the publications authored on each topic in the course of their career. 
    \item \textbf{T3:} Communicate the contribution of an author to each topic.
    \item \textbf{T4:} Illustrate the topics that two researchers have (or have not) in common, and how their contribution has been to each of the subfields.
\end{itemize}

\begin{figure}[ht]
    \centering
    \includegraphics[width=\linewidth]{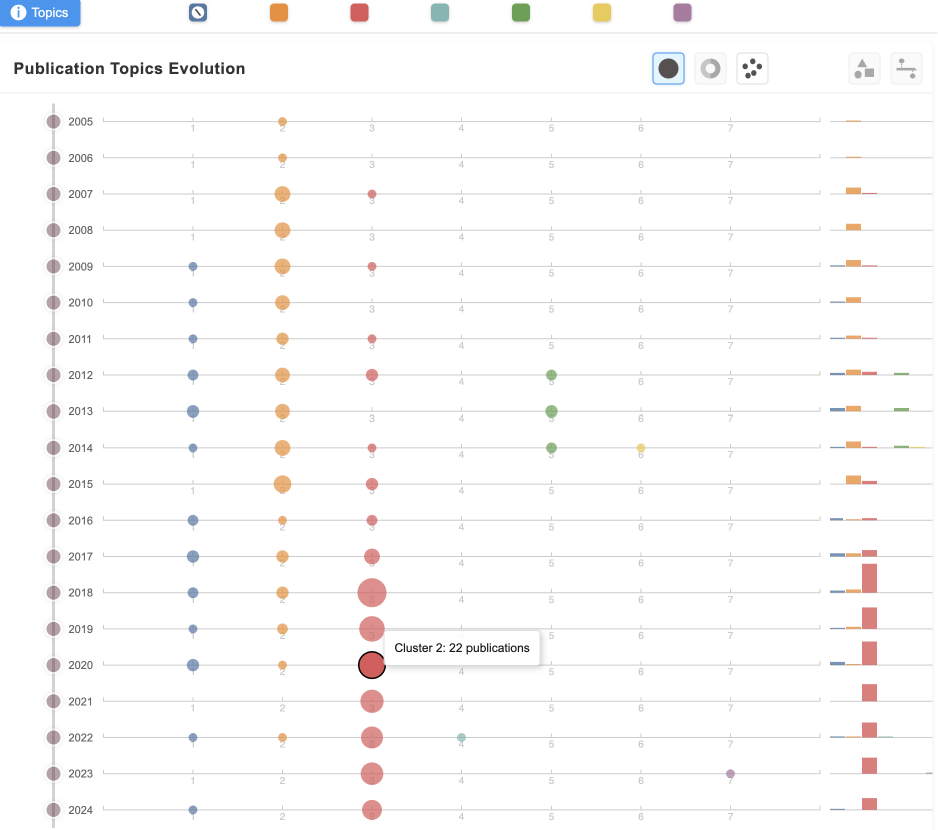}
    \caption{Evolution of research topics over the last 15 years of a researcher. It is clearly visible how the interest has shifted from Rendering (orange) to Deep Learning (red), and the number of papers published per year has increased (see red bar in the bar chart) recently.}
    \label{Fig: Researcher evolution}
\end{figure}

\begin{figure}[b!]
    
    \centering
    \includegraphics[width=0.8\linewidth]{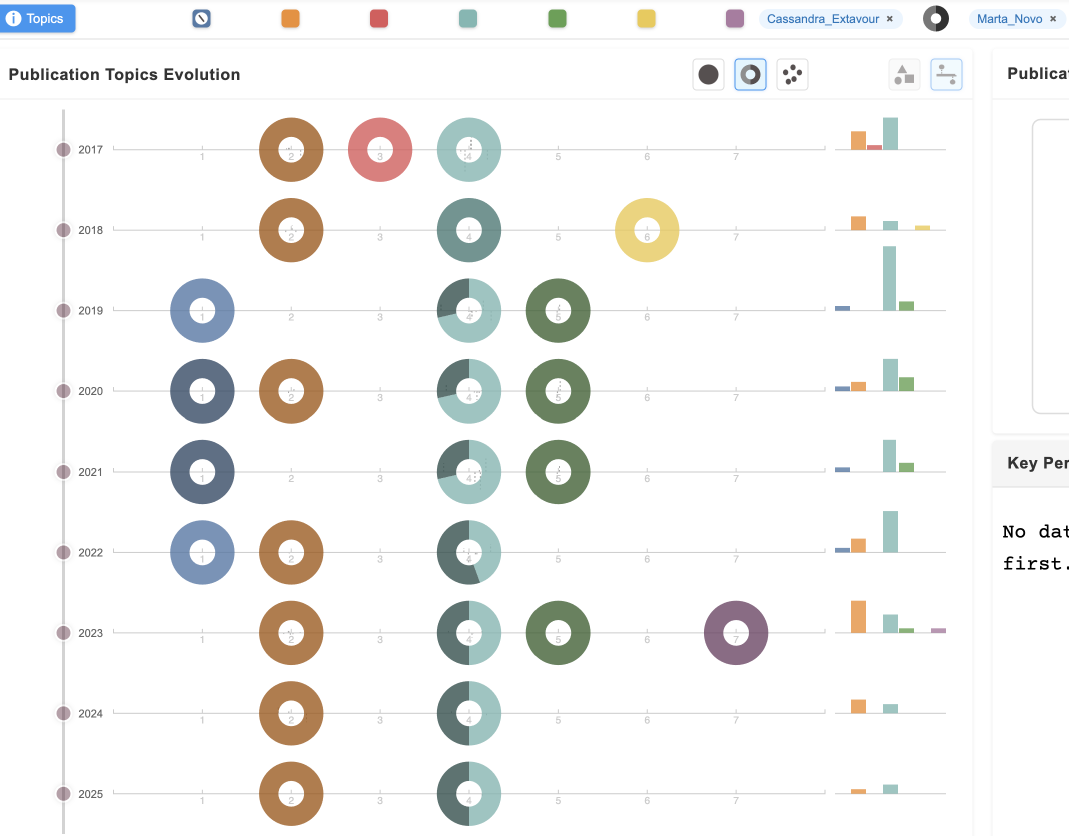}
    \caption{Comparing two researchers in the main view. The donut charts facilitate understanding of the distribution of publications between the first (lighter) and second (darker) researchers. These tones are used for all the clusters. Here only the last nine years are selected instead of the whole available time range.}
    \label{Fig: Comparing two researchers}
\end{figure}

Solving those tasks with a single visualization is not straightforward. Variations of commonly used techniques such as line charts, Priestley charts, or streamgraphs cannot easily communicate quantity, change over time, and facilitate comparisons at the same time. Since the two key elements to encode are topics and time, we select the two most significant visual variables to encode them: X and Y positions, respectively. And we interactively modify the marks at these positions to facilitate answering the indicated questions. Tasks \textbf{T1} and \textbf{T3} can be solved simultaneously by using a graduated circle at the corresponding positions, with the size indicating the quantity, as depicted in Figure~\ref{Fig: Researcher evolution}. Clusters are depicted both using position and color to \textit{facilitate their identification in other views}. The comparison between two researchers can be achieved by substituting the circles with donut charts with two different intensities of the cluster color, one per each researcher. This is illustrated in Figure~\ref{Fig: Comparing two researchers}. The light/dark encoding is depicted at the top with a small legend when this option is enabled, and, by hovering on the topic legend, the actual colors for each researcher are also shown.

Finally, the details of which papers were published each year can be solved by substituting the glyph with a small scatterplot per year and topic. This serves two objectives: providing individual marks so that the user can see the details of each paper (by hovering), and communicating the distribution of the papers within the cluster per each year, similar to information communicated by the 2D scatterplot on the top-right, but segmenting by time. The positions are based on the 2D coordinates of the top-right scatterplot but recalculated following an algorithm described in the supplementary material. In addition, we also added a small bar chart on the right to better communicate the distribution of the papers per topic each year. This is especially necessary when using the donut or the small scatterplot representations. The list of topics where the author has published can be obtained by clicking on the top-left blue button, which makes a panel roll out and shows the different topics the selected author(s) is involved in.

This design lets the user easily understand the topic evolution over time in a comprehensible overview, thus solving \textbf{GI2}. The combination of circles and donut charts lets users estimate quantities and compare the contributions of two authors, as required by \textbf{CMP1}.  

\begin{figure}
    \centering
    \includegraphics[width=0.8\linewidth]{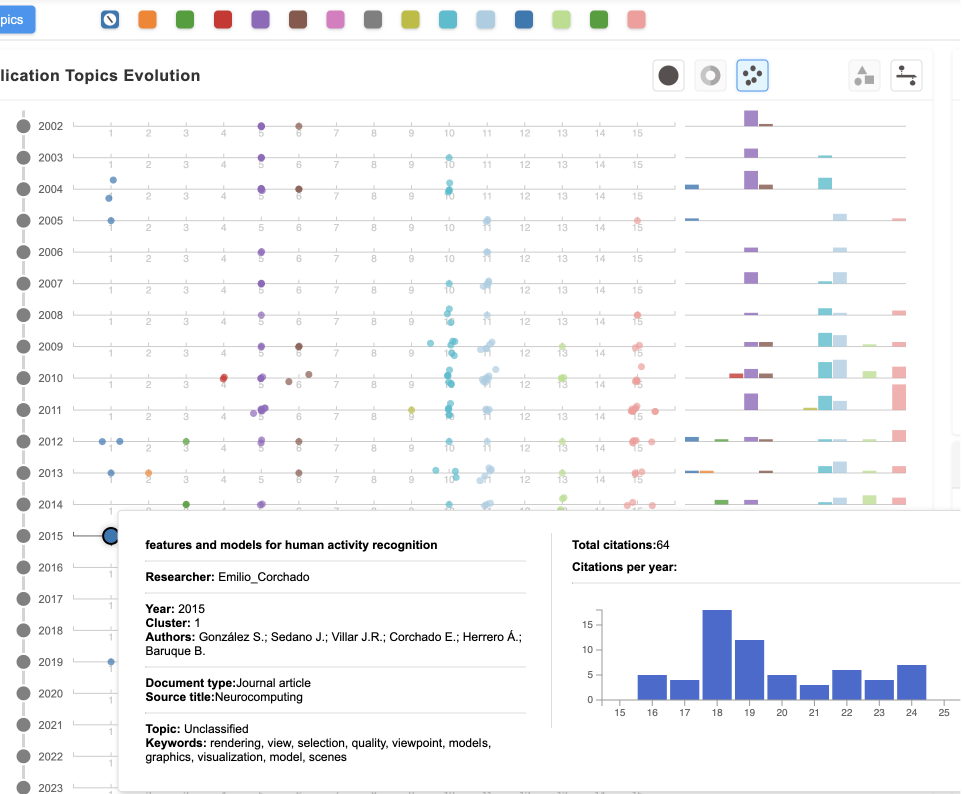}
    \caption{Points visualization with a hover operation that shows details on the hovered point.}
    \label{Fig: Points detail}
\end{figure}

\textbf{Interactions.} The change between the different modes is achieved through the buttons on top. To display two researchers, the option \emph{Multiple Researchers} in the header needs to be activated first. To further support detailed analysis, the users can select years by clicking or dragging on the year bullets, and upon clicking on the \emph{Apply} button, the years not included in the selection will be filtered out. The process can be set to default using the \emph{Reset} button, and the different steps of the selection can be rolled back with the \emph{Undo} button. Selections can be performed on all available visual depictions (circles, donuts, points, scatterplot, bars). As the outcome, we end up with a subset of publications that can be used to restrict the space of articles to be included in the prompt. On hover, these elements will show how many publications are selected (also with their authors, if necessary). \emph{Clicking} shows an extended details view of the publication that also contains the paper abstract. A lasso operator can also be used to select groups of visual depictions. In the \emph{points} view, the individual points can be hovered upon to obtain the details about the corresponding publication, and we can make the points to have a different shape per author. 

\begin{figure}[ht]
    
    \centering
    \includegraphics[width=0.8\linewidth]{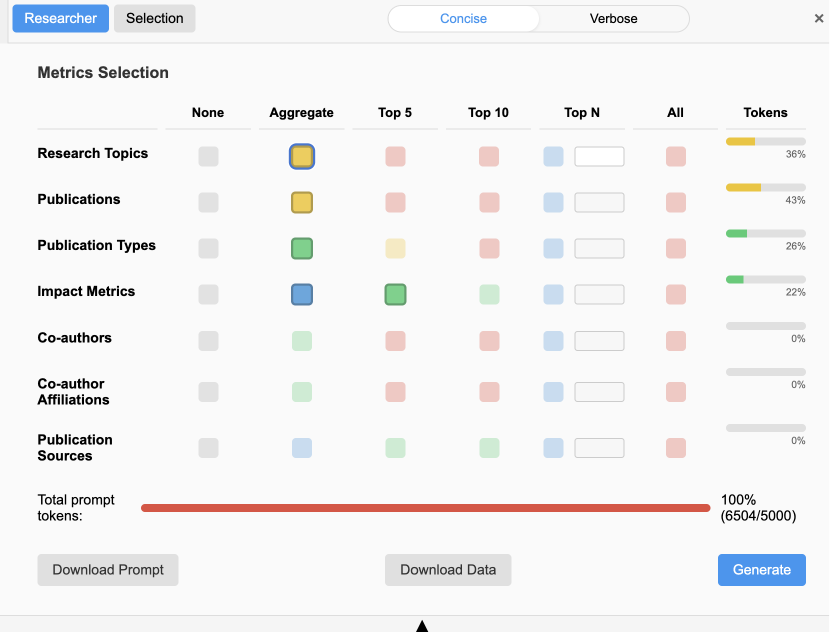}
    \caption{The prompt configuration tool lets the user indicate the metrics that will be present in the report, and automatically calculates the number of tokens (with a threshold of 5K) to ensure the limit is not surpassed.}
    \label{Fig: Prompt generation}
\end{figure}

\paragraph*{Report Generation}

To deliver a narrative description of the whole researcher's career highlighting the scientist's achievements (requirement \textbf{GI4}), our tool provides a descriptive report generator. This report can be configured to include the desired relevant insights (e.g., citations or collaborators), derived from the data and metadata of the articles, at different levels of aggregation. To do so, we have designed a visual tool where the user can easily and intuitively configure the relevant facts that must be included, as shown in Figure~\ref{Fig: Prompt generation}. Upon clicking the \textit{Generate} button, the application creates a prompt that includes the selected metrics, and instructs the LLM to generate a report based on this data. The tool also estimates the number of tokens the prompt will produce. Thus, the user is informed on how the inclusion of different parameters affects the output. This way, they can easily avoid exceeding the maximum number of tokens or calculate the price if using a non-free service.   

The prompt is then routed to an LLM using the Groq platform. Groq is a company that builds inference chips. To showcase their performance, the website has an interactive playground where the user can issue their prompts and offers a free API key with some usage limits (6000 tokens per prompt output). In our case, we used Qwen QwQ-32B~\cite{Yang:2024:Qwen2}, as it is the best-performing model at intelligence tests from the freely available through the GROQ API~\url{https://artificialanalysis.ai/providers/groq}. However, the choice of LLM can be easily changed in the code.

The prompt will look like this:
{\small
\begin{verbatim}
Generate a concise research profile analysis for <author> based on 
the following publication data:
## Research Topics Details:
The following lists detailed information about the researcher's all 
research topics. Analyze topic evolution, interconnections, and 
distinctive contributions to each area.
<data>
Instructions for analysis:
1. Analyze publication patterns and trends over time
2. Identify key research areas and their evolution
3. Evaluate citation impact and research influence
4. Assess collaboration patterns and networks
5. Summarize publication strategy (journals vs conferences)

Provide a focused, succinct analysis of approximately 150-200 words
highlighting the most significant patterns and insights. 
\end{verbatim}
}

The input data provided by the application in \emph{$<$data$>$} will be the set of publications if no previous selection was made, or a set of articles if the author selected them (in which case, the tool would highlight this condition). Selections are made by clicking on any visual depiction of the main view, as described before, or using the lasso selection on the dots or the scatterplot. The data injected to the prompt will be calculated based on the level of aggregation the user selects. Our application also fills the prompt with the name of the author.

\begin{figure}[h]
    \centering
    \includegraphics[width=\linewidth]{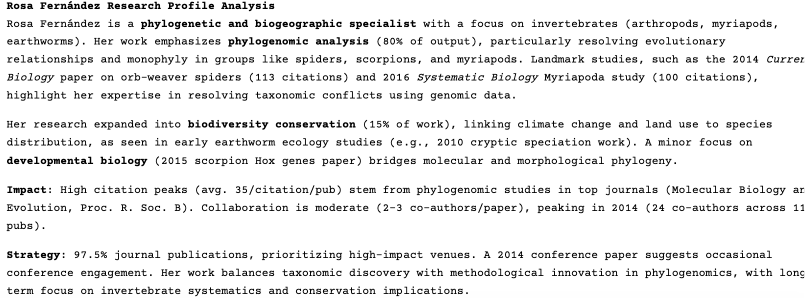}
    \caption{The concise report of the biologist Rosa Fernández, captured from the application. It was configured to show data on research topics, publications, publication types, and impact metrics.}
    \label{Fig: Report example}
\end{figure}

An example of the result of the prompt for a biologist named Rosa Fernández, taken from the application, is shown in Figure~\ref{Fig: Report example}. It was obtained by configuring the prompt to be \emph{concise}, and to deliver aggregated information about the research topics, publications, publication types, and impact metrics. In addition, for impact metrics and research topics, we asked for the top five elements to be included.

\begin{figure*}[ht]
    
    \centering
    \includegraphics[width=\linewidth]{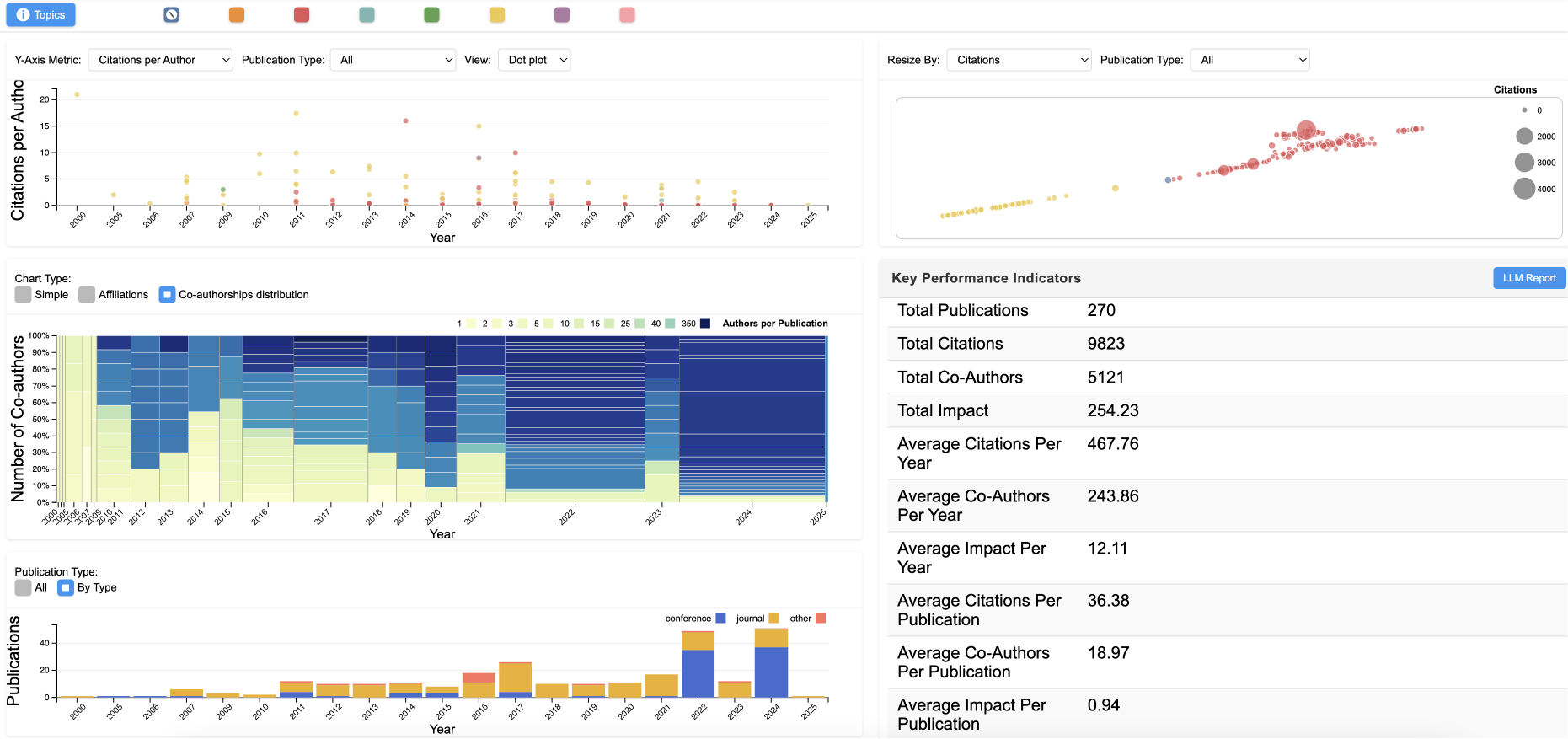}
    \caption{The analysis view displays several scientometric indicators for a researcher. The top-left view shows all the papers, and the Y axis can encode authors, citations, etc. The central view, serves to analyze the co-authorship, shown as bar charts, or as a Marimekko. The bottom view shows the types of publications. The bubble chart on the right shows the distribution of the papers, and the size of the bubbles can encode other scientometrics. Bottom-right, we can see some key indicators of the researcher in \emph{verbose} mode.}
    \label{Fig: InsightsView}
\end{figure*}

\subsection{Researcher Insights}

To obtain more insights on the career of a researcher, we gather data regarding a good deal of scientometrics. 
In assessment, promotion, or recruitment procedures, universities and evaluation institutions use metrics such as the total number of citations and the number of journal articles, and calculate scores based on elements such as the impact factor of the articles. To compensate for how different areas behave, they may use normalization factors per area (e.g., SNIP~\cite{Roldan:2019:Current}), or divide the impact factor according to the number of authors of each article. We have created a set of charts that transpose such elements employing the SJR impact factor, the SNIP relative impact factor, and by providing normalization based on the number of authors of a publication upon user request. 

Note that the use of the number of citations is prone to all sorts of misinterpretations. It is well known that the number of references may vary significantly between areas of research. Furthermore, the relevance that citation count has been given lately by recruiting or promotion processes at academic institutions has led to many sorts of malpractice~\cite{Joshi:2024:Deception}, including the creation of so-called "citation mills" (a type of citation-boosting service that creates fake citations to an author's work)~\cite{Ibrahim:2025:Citation}. Therefore, instead of Google Scholar citations counts that are easy to manipulate, we rely on Scopus metrics that use only peer-reviewed sources and citations to peer-reviewed articles for their calculation~\cite{Roldan:2019:Current}.

These insights are showcased as a set of charts that focus on time development, as shown in Figure~\ref{Fig: InsightsView}, and are intended to solve the Personal Insights requirements \textbf{PIM2} to \textbf{PIM5}. The initial configuration serves as overview of the data, centered in citations, coauthors, and number of papers, over time.
First, on the top-left, we have a dot plot that displays the papers published along a year. The X-axis encodes the years and the Y-axis encodes the metric we want to highlight, which can be changed by the user. By default, this metric is set to the number of citations of the papers. We can also highlight the average of citations with a line or, as we have done here, the number of citations per author. Below, we have the collaboration view. Initially, it shows a bar chart with the number of co-authors who have collaborated with the researcher every year. But we can change the chart to breakdown collaborators per affiliation (same, same country, and international), or display both the collaborators and the number of papers, using a Marimekko chart, as shown in Figure~\ref{Fig: InsightsView} (central part), where we can see numerous (actually hundreds if we hover over the rectangles) co-authors in the last years. At the bottom, another bar chart shows the number of publications per year (that can be broken down by publication type). 

The top-right view shows the distribution of the publications as a scatterplot. This can be modified to a bubble chart and the size of the bubbles can be dictated by the number of citations of the publication (as in the image), or the impact of the publication in Scimago ranking (SJR), as well as the normalized relative impact per area, the SNIP factor. The bottom right provides a concise set of indicators, that can be made more verbose (with the corresponding button) and filtered per period. Like in the previous view, we can also generate the report here.

\textbf{Interactions.} We have designed a set of interactions that let the user drill down into the different variables through many optics. The dot plot can encode on the Y axis the following parameters: citations, number of authors, citations per author, citations per impact, share of citations, or impact (SJR or SNIP). This quantity is modified through the dropdown menu at the top-left. In addition, the publications can be filtered to journals, conferences, or others. The user can also show a line with the average or the sum of the Y-encoded metric. As already mentioned, the center-left chart can represent the number of co-authors, disaggregate them per affiliation, or show the yearly detail of co-authorship through a Marimekko chart that has the number of articles encoded as the width of the bars and the different numbers of co-authors in the different levels of the stacked bar. The bottom chart can split the values by publication types. Additionally, all visual elements provide details on hover and the publications (dots) show more detailed contents on \emph{click}. The scatterplot can be turned to a bubble chart by selecting the resizing variable (citations, co-authors, $\ldots$) on top of it.

\subsection{Comparative Time-Based Visualization}

One of the most common usages for the data extracted from Scopus is the comparison of different researchers. For competitive procedures, such as applications for grants or positions, many of the variables we gather or calculate are commonly employed to compare different candidates. Note that several metrics, such as impact factor, number of collaborators, and citations, are highly dependent upon the research field. We envision the scenario where a set of candidates is being considered for a commission or evaluated for an open position. In this case, it is logical to presume that the candidates would belong to the same or very close areas. In this situation, it may make sense to compare metrics such as citations or impact over the years. Hence, we have designed a third tab intended to solve the tasks \textbf{CMP2} and \textbf{CMP3}. In this case, instead of focusing on the comparison of two authors, we facilitate comparisons of three or more researchers. Although including more than five researchers makes the charts difficult to read, and the rightmost KPI table will require horizontal scrolling. The comparisons are operated through a series of line charts, with the same set of variables as the base analytics view, but enabling multiple authors at once. 

\textbf{Interactions.} The line charts in this view, in addition to researcher selection, also provide the same metric changes previously described. They also show details of the values on hover. This, combined with the \emph{verbose} version of the KPI view, enables gathering detailed insights on the scientists.

\begin{figure}
    \centering
    \includegraphics[width=\linewidth]{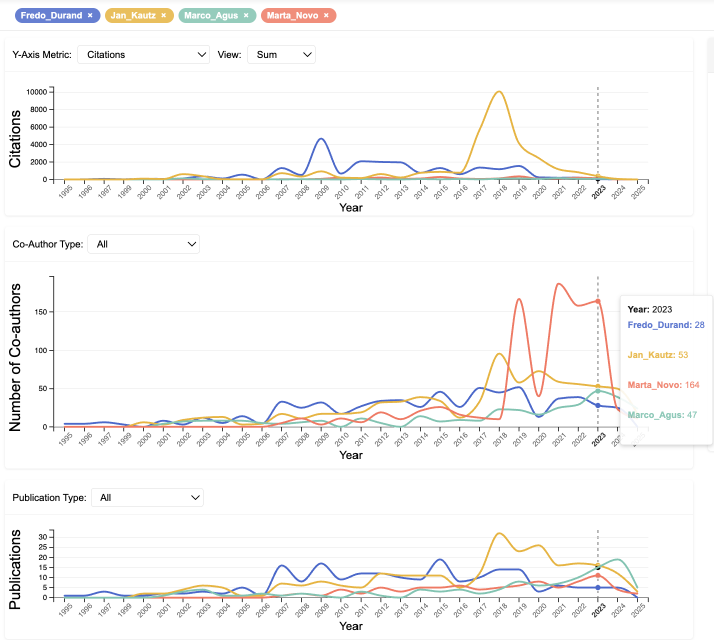}
    \caption{Multiple researchers' analysis. The analysis views allow for a temporal comparison of multiple researchers at once. The metrics can be changed interactively, as well as the researchers in consideration. The right panel (here clipped) contains the Key Performing Indicators of the selected researchers.}
    \label{fig:Panel2MultipleResearchers}
\end{figure}

\section{Application scenarios}
To demonstrate the usefulness and applicability of our proposed solution, in the following, we sketch five exemplary scenarios that cover the most typical usages that we envision. 

\paragraph*{Scenario 1: Understanding the Topics of Interest and Progress.}
In our academic life, we often need to determine the topics of interest of a scientist and how they have evolved. For example, when searching for colleagues to apply for a project or when a student is looking for a Ph.D. supervisor. 

\begin{figure}[ht]
    \centering
    \includegraphics[width=\linewidth]{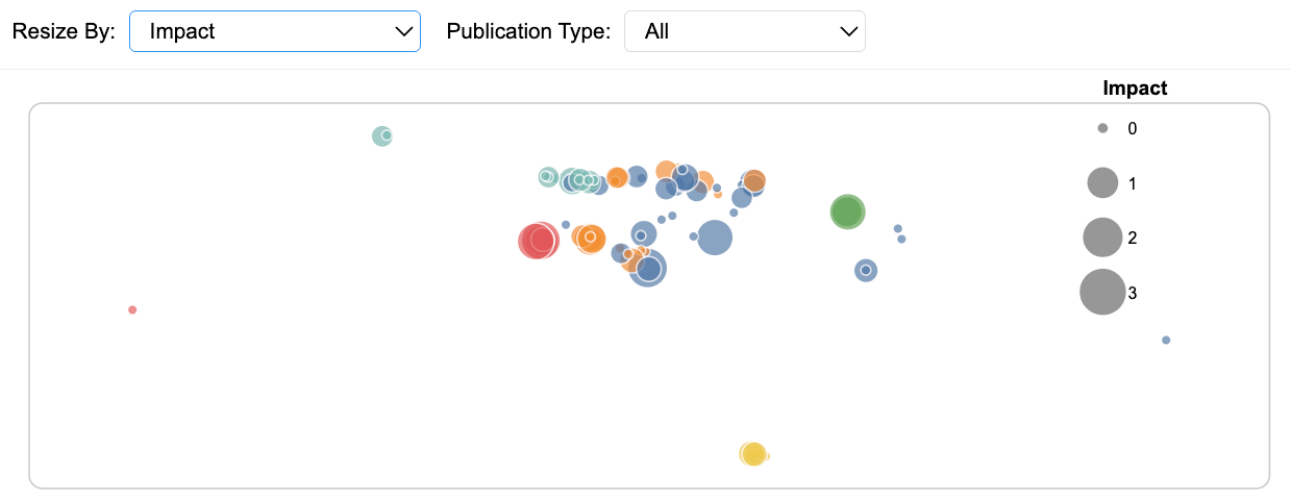}
    \caption{Assessment of the relative impact of publications of the same author by using a bubble chart that encodes the relative impact as the size of the bubbles.}
    \label{fig: Impact Scenario 1 - Smart Cities}
\end{figure}

Let's imagine we are considering a person as a supervisor in the area of Smart Cities. We may open the main view of a researcher (Figure~\ref{Fig: Researcher evolution}). We can quickly check their topics of interest using the \emph{Topics} list. In this case the yellow cluster includes the keywords Smart Cities and Internet Things.
This indicates contributions in this area, which appear to be recent enough and growing. Turning to the analytics view, we can determine how important those contributions are for the researchers' profile. The top-right scatterplot shows the publications for the selected years. These publications are easy to identify since they are all far from the rest of the publications of the researcher. To determine how relevant those publications are compared to the other publications of the researcher, we can apply a resizing to the scatterplot that depends on the impact, as shown in Figure~\ref{fig: Impact Scenario 1 - Smart Cities}. The user can also generate a report on these selected publications to get more detailed insights.

\paragraph*{Scenario 2: Comparing Two Scientists}
Researchers oftentimes need to find collaborators for applying for a project, whose expertise is complementary to theirs. Or, when we have two candidates for a role in an evaluation board, we may need to determine who has contributed more significantly to the field where both are publishing. \textbf{VizCV} enables direct comparison to solve these tasks. After opening the \textit{Overview} and selecting the authors to compare, the user can choose the donut chart mode. To better examine the recent research, the user can select a period (e.g., the last 9 years), on the left and apply it to filter out the other years, as shown in Figure~\ref{Fig: Comparing two researchers}. In this case, we see that the two authors are biologists, but share only one area \emph{developmental biology}. In the remaining areas, they are complementary.

\paragraph*{Scenario 3: Assessing Expertise of a Candidate for a Commission or as Evaluator}
Research institutions often require finding adequate profiles to fill positions in regulatory committees or evaluation boards (e.g., to act as Ph.D. jury). In such cases, it may become necessary to assess the expertise of a potential candidate. 

This process can be easily solved using \textbf{VizCV}, as follows. Let's imagine that we are considering one researcher for a position on a commission for evaluating Virtual Reality project applications. To assess their expertise, we can go to the main view, load their profile, and check what are their areas of expertise. As a research project proposal evaluator, we would expect a long-term experience of the researcher. We can quickly inspect the topics and see whether VR or related areas are present. Once we have done so, if a fine-grained exploration is required, we can swap to the \emph{dots} view and hover over the individual articles to see their titles. By clicking on the dots, the abstract is also accessible. This allows us to get more detailed information. We can also check the impact of those publications by swapping to the \emph{Insights} view and configuring the dot plot so that citations (or impact) are used on the Y-axis. Furthermore, we can select the subset of papers we are interested in and generate a survey using the report generation tool. This allows us to have written form of the main features that justify their selection.

\paragraph*{Scenario 4: Performance Evaluation}

In academia, there are many occasions in which the performance of a researcher needs to be evaluated. \textbf{VizCV} can support evaluation procedures by helping the users to gain detailed insights on indicators, such as the number of journal papers, the number of citations, the impact factor, as well as the impact factor normalized by the number of authors. These metrics are all used in the universities of the authors to assess the profiles of candidates for promotion processes, for example. 

To provide supporting information from a candidate, we can follow the next steps. First, the user can open the main view of the researcher and create an individual report that describes the most relevant aspects of their publications track record, which would look like as an extended version of the one shown in Figure~\ref{Fig: Report example}. Then, by turning to the \emph{Analytics} view, the user can analyze the relevant Key Performance Indicators and further drill down into the data. For instance, in Figure~\ref{Fig: InsightsView} we can see an author who has a high number of publications in two disparate areas: neutrino astronomy and underwater acoustics. When analyzing the citations, more than 9000 in this case, one clearly sees that most of them come from physics articles. Indeed, a single paper (the big red bubble) is responsible for more than 3000 citations. In addition, as shown in the Marimekko chart on the left, many of these papers have hundreds of authors. By displaying the citations per author in the top-left chart, one can see that the underwater acoustics (yellow) have a higher proportion of citations than physics ones. Another noticeable observation is that the number of publications, mostly journals, has increased significantly over the last years, with around 50 publications in 2022 and 2024. This is easily seen by checking the bottom left bar chart. All this information can be used to support a broader evaluation of a profile for promotion.

\paragraph*{Scenario 5: Recruiting}

For the recruitment of talents, the recruiting commission commonly needs to evaluate more than a single profile. For an academic position, commonly the profile will be determined by the call, although it might be slightly open. However, the candidates will likely have similar profiles. In this case, the direct comparison of all candidates with scientometrics may make much more sense. Our tool can be used to first generate a descriptive account of the candidates' profiles using the report generator with the desired indicators. Then, accurate comparisons between multiple researchers can be performed using the Analytics view with the Multiple Researchers option activated, as shown in Figure~\ref{fig:Panel2MultipleResearchers}. In this case, a set of line charts can be fine-tuned to show different aspects of the publications tracks: number of citations per year (top), number of collaborators (middle), and the publications per year (bottom). To get more insights, KPIs are also measured per researcher and displayed in the rightmost view. Those insights can be added to the reports if desired and may complement other information the evaluators may need to consider for the process. Note that this report can be restricted to a subset of publications.

\section{Evaluation}

We have assessed the potential of our tool within a set of semi-structured interviews with three university officials. The first one is the current adjunct to the Personnel vice-rector. Together with the Personnel vice-rector, he is responsible for defining the procedures, rules, and relevant metrics for recruitment to faculty positions. The second person is a former head of the university board for evaluation and selection of personnel. This board is responsible for the evaluation of faculty members for promotion, or when they apply for a sabbatical period. The third person is a vice-rector for Digital Strategy (and former head of a research group), who supervises the maintenance and creation of new tools of governance at the university. 

The evaluation procedure was as follows: we showed the participants the tool, explained how the topic classification works, and how it was extracted from the data. We also depicted the different features of the tool in an interactive session and then showed a demonstration of how the tool may help in solving individual scenarios presented in the previous section. After a discussion on the features and limitations of the tool, we asked them whether the scenarios were realistic at their university, whether the tool was suitable to solve the described scenarios, and whether they would prefer using the tool over existing ones or their traditional workflow when solving these tasks. They all agreed that the developed tool was very useful when dealing with all those scenarios. They all said that these scenarios were realistic and that they would use the tool to help them if such a tool was available. They also said that the available tools of the university were either not solving (scenarios 1 and 2) or only partially helping (scenarios 3-5) to solve the tasks, but in those cases, they required a lot of time to use and analyze the data that was provided. The former head of the selection board praised the tool and said it would have been highly useful for the board. However, for some procedures, additional information is needed, such as the gender or the actual position of authors in the publication (if relevant). They also pointed to other use cases, such as the evaluation of candidates in project applications. They also commented that the tool can be especially suitable for highly competitive recruitment calls because it provides them with fine-grained details about the candidates. They also suggested adding new features, such as a normalization of the citations per topic or a visual comparison of KPIs.

\section{Discussion}

Our application leverages several cutting-edge together with other classical, AI-based technologies, to increase the expressivity of our system: We detect topics using BERTopic~\cite{Grootendorst:2022:Bertopic}, embeddings are generated with Specter~\cite{Cohan:2020:Specter}, the layout and the clustering makes use of UMAP~\cite{McInnes:2018:UMAP} to reduce dimensionality, and HDBSCAN~\cite{McInnes:2017:HDBSCAN} to detect clusters. Finally, the different textual reports are generated using LLMs (the default is Qwen QwQ-32B~\cite{Yang:2024:Qwen2}) with customized prompts injected with relevant KPIs of the researchers.

As a result, our system has several \textbf{advantages} over other systems: \textit{a)} It works with a relatively low resources setup since the input data can be obtained through Scopus, and many universities have agreements with Elsevier that allow researchers to download Scopus profiles and obtain API keys at no extra cost. \textit{b)} It provides a rich exploratory visualization of one or multiple researchers. \textit{c)} The text generation system works with a free API key from Groq.com. \textit{d)} It is capable of answering multiple questions that range from a holistic global view of individuals and specific metric-constrained reports of researchers to detailed insights that can be used to compare multiple researchers. \textit{e)}. We set the temperature of the system to 0 so that the report precisely analyzes the KPIs we inject into the prompt. Moreover, our system scales seamlessly to a full research group or department if considered as a single entity.

Our approach also has \textbf{limitations} regarding the data and the algorithms used. 
For the paper representation and topic modeling, we used only information on the abstracts and index keywords. Our pipeline can also work with full papers, but we found it difficult to automatically download the full set of authors' publications. Even when they are publicly accessible, the scrapers we wrote encountered issues, such as being blocked by the host server. Our current implementation algorithm works reasonably well, but occasionally the articles are not classified where we would expect them. We plan more experiments to obtain a more robust automatic fine-tuning of the classification. Working with full papers would likely improve the topic selection. On the other hand, Scopus profiles are easily accessible. Moreover, we envision our tool to be used by anybody just by uploading the CSV files of the Scopus profiles. The scripts to process the data can work in the background.  

The system scales well for a full research group represented as a single entity. But if many authors from many distant areas are uploaded, this might result in many topics that would not be easily showcased in the main view. A possible solution is to increase the threshold for the fusion of topics or make use of the BERTopic built-in topic reduction.

Some experts, who were shown a former prototype version, asked about the Google Scholar citations. Like the texts of full papers, we had also included them in the previous version of the application. Nevertheless, Google Scholar citation numbers are easy to game through \emph{citation mills} or preprint servers, as demonstrated previously~\cite{Ibrahim:2025:Citation}. Thus, we preferred to stick to more robust metrics. But these metrics could be simply added or can substitute the current metrics we use. 

\section{Conclusions and Future Work}

In this paper, we have presented \textbf{VizCV}, an end-to-end pipeline for data extraction and processing of researchers' profiles from Scopus data, and a web-based tool for the exploratory analysis of researchers' publication track records. \textbf{VizCV} can be used to analyze a single or multiple researchers' careers, to assess their expertise, to compare their scholarly evolution, or even to support the evaluation for promotion or recruitment. Our tool has been examined by three university officials highly familiar with evaluation, promotion, and recruitment procedures. They found that the tool could be very useful for the described scenarios. In the future, we plan to add an option to upload a profile of a desired author. But the duration of the whole process is inevitably upper-bounded by document embedding creation, dimensionality reduction, and topic modeling, which take up to one to two minutes for almost 40 authors with careers of 20+ years. This limitation could be addressed by making the process stages configurable and splitting the waiting times to ensure a smoother user experience. In the future, we plan to add a button to upload the desired author profile. We also want to add other metrics (e.g., Web of Science citations) and analyze other methods to provide insights into the researchers' impact. Moreover, we would like to enable an option to import researchers' data directly from the universities' internal databases. We also plan to publish the code repository if the paper is accepted.

As supplementary material, we provide a video showcasing the usage of the tool, the re-projection algorithm for the mini-scatterplots in the dots view, and extra examples of prompts and the output of the LLM.

\ignore{

\acknowledgments{%
	The authors wish to thank A, B, and C.
  This work was supported in part by a grant from XYZ (\# 12345-67890).%
}
}

\bibliographystyle{abbrv-doi-hyperref}

\bibliography{bibliography}

\end{document}